# Design of Chemical Propellant Thruster to Deorbit Nano satellite: StudSat-II

**Prabin Sherpaili**
Research Scholar
Centre for Small Satellite Development
Nitte Meenakshi Institute of Technology, Bengaluru, India
Email: sherpailiprabin007@gmail.com

**Roshan Sah**
Space Robotics/Propulsion Researcher
Space Robotics Lab
Tata Consultancy Services (TCS) – Research
Bangalore, India
Email:- sah.roshan@tcs.com

**Sandesh Hegde**
Research Scholar
Centre of small satellite development,
Nitte Meenakshi Institute of Technology, Bengaluru, India
Email:sandesh@teamstudsat.in

**Bir Bahadur Chaudhary**
Research Scholar
Department of Mechanical and Aerospace Engineering
Pulchowk Campus, IOE, TU, Lalitpur, Nepal
Email: 076msmde004.bir@pcampus.edu.np

**ABSTRACT**

*An increase in the application of the satellite has skyrocketed the number of satellites, especially in the low earth orbit (LEO). The major concern today is after end-of-life, these satellites become debris which negatively affects the space environment. As per the International Guidelines of the European Space Agency, it is mandatory to deorbit the satellite within 25years of end-of-life. This paper is aimed to design the solid chemical propellant thruster to deorbit the StudSat-2 from its original orbit to the lower orbit. StudSat-2 carries the heritage of StudSat-1 which was successfully launched on 12th July 2010 AD and is the first Pico-Satellite in India by the undergraduate students of seven engineering colleges. This paper explains how a solid monopropellant thruster could be used to deorbit the satellite after the end-of-life with the least difficulty compared to other active and passive methods of deorbiting. The deorbiting mechanism consists of a solid propellant, Convergent-Divergent nozzle, ignition system, and electronic actuators. The components of the thruster were designed in the CATIA V5, and the combustion studies and flow analysis were done in ANSYS. The concept of Hohmann transfer was used to deorbit the satellite and STK was used to simulate it.*

**NOMENCLATURE**

| | | | |
|---|---|---|---|
| $V_t$ | Velocity in transfer orbit | P | Pressure |
| $\mu$ | Gravitational parameter for Earth | T | Temperature |
| $R_1$ | Radius of initial orbit | M | Mach number |
| $R_2$ | Radius of final orbit | V | Velocity |
| $\varepsilon_{transfer}$ | Specific mechanical energy in transfer orbit | M | Mass |
| $a_{transfer}$ | Major axis of the transfer orbit | $U_{eq}$ | Equivalent velocity |
| $\varepsilon_{orbit}$ | Specific mechanical energy in orbit | $\varepsilon$ | Expansion ratio |
| $\Delta V_1$ | Change in velocity for first transfer | d | Diameter |
| $\Delta V_2$ | Change in velocity for second transfer | A | Area |
| $\Delta V$ | Total change in velocity for complete transfer | t | Thickness |

## 1. INTRODUCTION

After the launch of the sputnik 1, the space race officially begun. Technological developments and commercial purpose have significantly increased number of satellites in Low Earth Orbit. After end of life, these satellites poise threat to functional satellites (catastrophic collision) as well as occupy precious slots. To mitigate this issue, Inter-Agency Space Debries Coordination Committee have formulated guidelines, which includes mandatory re-entry of satellite after 25 years of end-of-life.

American and Soviet missions have performed various re-entery missions to recover valuable hardware data, satellite (Yantar-2K) or human spaceflight capsule (eg. Gemini flights). Project STUDSAT-2 [1] takes heritage from STUDSAT-1. It is India's First Twin Nano Satellite Mission, which aims to demonstrate Inter-Satellite Communication and De-orbit satellite after mission life. Several deorbiting mechanism were studies and solid propellant thruster is most studied one. Solid propellant thruster is used to deorbit StudSat-2 from its initial orbit to lower orbit where it burns off due to atmosphere.

Even though the space robotic concept has been continuously evolving which is recently done by Roshan et. al. [2][3][7][8][9], where a allocated spacecraft is used in which a UR5 robotics manipulator [6] [16] is used for the space debris removal whose main function is to chase, capture and de-orbit the debris object or spent satellite to very low earth orbit. The orbit transfer [4], and space rendezvous and docking become one of the important factors to reach close to the space debris object. This kind of spacecraft/satellite also used the RCS thruster which is the combination of chemical thruster along with the attitude control sensor [5] for better operating of the thruster in three directions.

## 2. METHODOLOGY

Hohmann method is used to Deorbit StudSat-2 [1], which consists of initial, transfer and final orbits. Hohmann transfer is an orbit transfer maneuver between two orbits of different altitude around the central mass with the elliptical orbit. For deorbit, it requires two impulsive burn: first establishes elliptical transfer orbit and second adjusts orbit to make final circular orbit. Propulsion system used in the project is solid motor propellant and consists of four thrusters, two oppositely placed thrusters are fired at a time. Timing of burns are critical to determine the orbit of deorbit. So, it is determined by micro-controller analyzing data from the sun sensor and magnetometer for position.

From the initial and final orbits, delta velocity (braking velocity) is calculated. Thruster is designed considering orbital parameters and fuel properties, which would provide required propulsive force for deorbit. Thrusters are modeled in CATIA V5 and flow simulation is performed in ANSYS. To demonstrate the orbit transfer [4], System Tool Kit is used which gives altitude change during the processes.

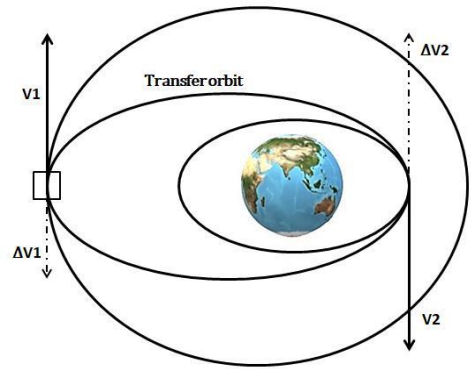

Fig.1: Hohmann deorbit mechanism.

## 3. DEORBIT MECHANISM:

The mission of de-orbiting using solid propellant thruster is to bring the satellite from initial low Earth Orbit of 600 km to 200 km [8][9]. This is achieved by first transferring the satellite in the transfer orbit whose apogee is 600 km and perigee is 180 km. This transfer will be accomplished by firing the thruster. When the satellite reaches perigee again the thruster is fired to transfer it in the final circular orbit of 200 km. The satellite is de-orbited to 200 km altitude where it encounters a high atmospheric drag. This drag finally leads to the complete disintegration of the satellite due to friction and heat generated over the surface of satellite. The necessary delta-v for descending of the satellite to the final altitude is provided by the solid propellant thruster. The thrusters are placed on the front face of the satellite that will produce the delta-v required to decrease the total energy associated with the satellite, hence decreasing the altitude of the satellite.

## 3.1 Orbit Determination:
Velocity of the satellite at the perigee of the satellite

$$V = \sqrt{2\left(\left(\frac{\mu}{R_{orbit}}\right) + \varepsilon_{transfer}\right)} \quad (1)$$

$$\varepsilon_{transfer} = -\mu/2a_{transfer} \quad (2)$$
$$2a_{transfer} = R_{orbital\,1} + R_{orbital\,2} \quad (3)$$

Delta velocity required for the transfer from initial orbit to apogee of transfer
$$\Delta V1 = |V_{orbit1} - V_{transfer\,orbit}| \quad (3)$$

Delta velocity required for the transfer from perigee of transfer orbit to final orbit
$$\Delta V2 = |V_{transfer\,orbit} - V_{orbit2}| \quad (4)$$

Total delta velocity for the transfer of orbit
$$\Delta V = \Delta V1 + \Delta V2 \quad (5)$$

| Parameters | Calculated value for 680 to 180 Km | Unit |
|---|---|---|
| R earth | 6370 | Km |
| R orbit1 | 6970 | Km |
| R orbit2 | 6570 | Km |
| a transfer | 6770 | Km |
| ε transfer | -29.44 | Km2/s2 |
| ΔV 1 | -0.1125 | Km/s |
| ΔV 2 | -0.1142 | Km/s |
| ΔV | -0.2268 | Km/s |

## 4. DESIGN OF PROPULSION SYSTEM:
### 4.1 Selection of propellant:
Various methods can be deployed to deorbit satellites [12] [13] after the end-of-life which include cold gas thrusters, liquid and solid propulsion [10], electric thrusters, teathers, drag sail etc. Low specific impulse, thrust limits is application only for orientation rather than deorbiting [14][15]. Electric thruster is complex with low thrust and higher power consumption. System complexity, toxic propellant and cost outweights the advantage of liquid propellant. Teathers and drag sail require long time for deorbit. Solid propellant is promising due to its simplicity, high-density performance and storability.

Bayern-Chemie from Germany have developed solid motor rocket for D-SAT mission. It uses particle free propellant (RESI REduce SIgnature composite propellant) which has excellent aging characteristics. It was artificially aged for 10.5 years at 62 °C and had similar properties as new propellant claiming shelf life of 12.5 to 16 years.

Key properties of Bayern-Chemie RESI propellant:
Burn rate            12-17 mm/s (Pc= 100bar at soak temperature 20 °C)
Burn rate exponent   0.25-0.5
Density              1700Kg/m$^3$
Specific impulse     2280-2420 m/s (expansion ratio 70:1)
Specific impulse     2700 m/s (vacuum)

### 4.2 Principle of operation:
The chemical energy of solid propellant (RESI composite propellant) is converted into the mechanical energy through aero-thermodynamic processes. These inturn reduces mechanical energy of the StudSat-2, and transfers it to lower orbit.

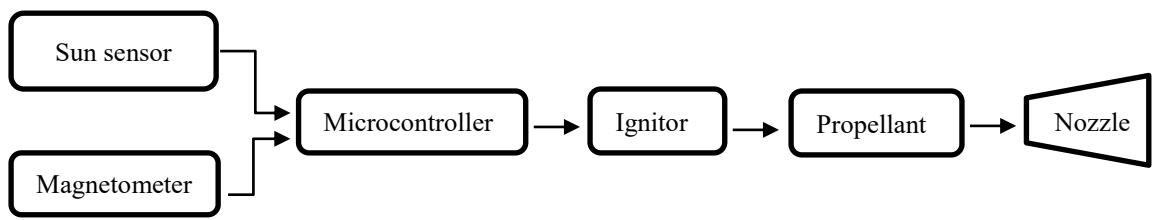

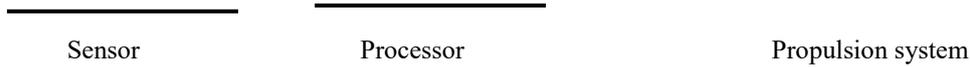

Sensor             Processor             Propulsion system

Fig.2: Schematic diagram of de-orbiting system.

### 4.3 Design of combustion chamber:

Mass of propellant required for the required ΔV is obtained using below mentioned formula

$$\frac{mi}{Mf} = e^{\frac{\Delta V}{ueq}} \qquad (6)$$

Aluminum combustion chamber [11] is designed to store the solid rocket motor. The insulator coating prevents transfer of heart to the chamber. To withstand high pressure during the combustion, thickness of chamber is determined by maintaining safety factor of 1.5 using thin cylinder concept.

$$.t = \frac{p*d}{2\sigma} \qquad (7)$$

Parameter:

| | | |
|---|---|---|
| Pressure inside combustion chamber | 7 | Mpa |
| Mass of propellant | 0.44 | kg |
| Number of thruster | 4 | mm |
| Length of combustion chamber | 60 | mm |
| Internal diameter of combustion chamber | 38.2 | mm |
| Thickness of combustion chamber | 2.11 | mm |

### 4.4 Design of convergent divergent nozzle:

After the combustion of solid motor propellant, hot gas is expanded through convergent divergent nozzle.

$$\frac{Pc}{Pe} = \left(1 + \frac{\gamma-1}{2} Me^2\right)^{\frac{\gamma}{\gamma-1}} \qquad (8)$$

$$\frac{Tc}{Te} = \left(1 + \frac{\gamma-1}{2} Me^2\right) \qquad (9)$$

$$\varepsilon = \frac{Ae}{At} = \frac{1}{Me} \sqrt{\left[\frac{2}{\gamma+1}\left(1 + \frac{\gamma-1}{2} Me^2\right)\right]^{\frac{\gamma+1}{\gamma-1}}} \qquad (10)$$

Parameter:

| | | |
|---|---|---|
| Diameter of throat | 6.35 | mm |
| Diameter of exit | 18.14 | mm |
| Length of nozzle | 75 | mm |
| Mach number at exit | 2.82 | |
| Diameter of inlet | 10 | mm |

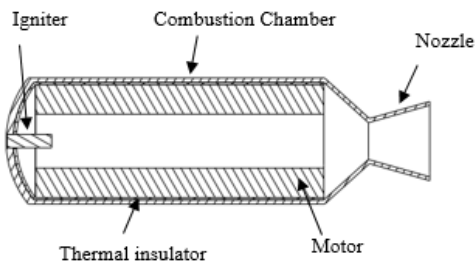

Fig.3: component of thruster.

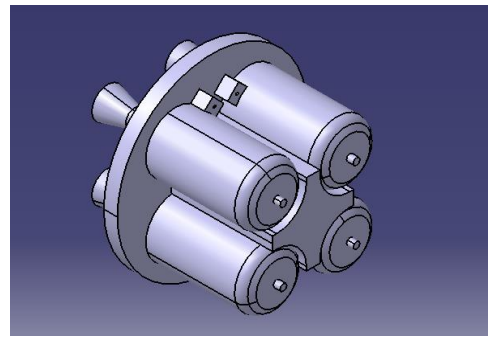

Fig.4: Propulsion System.

**4.1 Verification from Simulation:**
   For the simulation, standard material properties are considered. ANSYS is used to perform structural simulation of the combustion chamber. Geometry is modelled in CATIA and imported to ANSYS. Fine mesh is created with 80 thousand node points. The pressure load of 70bar is applied as it is the highest pressure in combustion chamber. For the simulation of the convergent divergent nozzle, k-omega turbulence model [17] is used. During meshing 10 lakhs node point are created. The inlet pressure is 50 bar and temperature is 3000K. The outlet pressure of 0.1 bar is considered making the flow under-expanded. Variation of static pressure, temperature, mach number, density etc is observed.
   For the orbital simulation, STK is used. The initial orbit of 600Km is created from where velocity change (ΔV) is given to transfer it to desired lower orbit of 200Km. Varation of Semi-major axis, inclination, RAAN, etc is observed.

## 5. RESULT:
**5.1 Propulsion System:**
Structural analysis of Combustion chamber is performed in Ansys [11]. The average von-mises stress of the chamber is below the ultimate yield strength. This shows combustion chamber is strong enough the bear high combustion pressure.

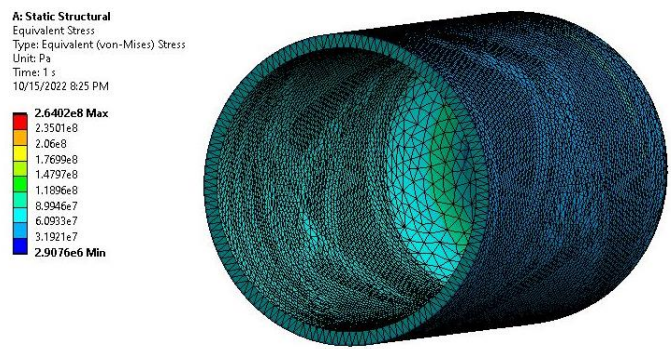

Fig.5: Von-mises stress distribution in combustion chamber.

With the help of ANSYS, following Mach number, pressure and temperature contour are obtained.

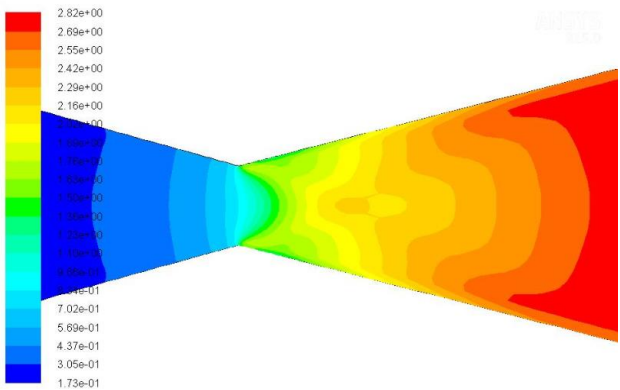 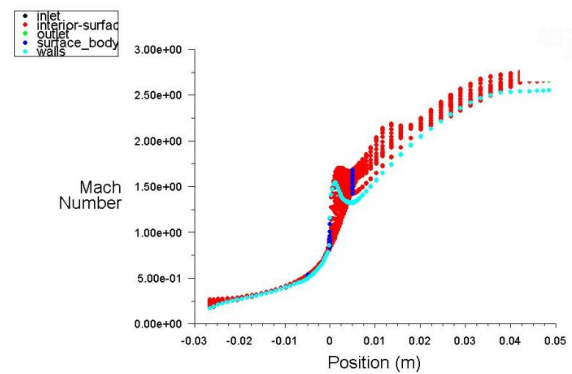

   Fig.6: Mach number contours .                               Fig.7: Mach number variation along nozzle.

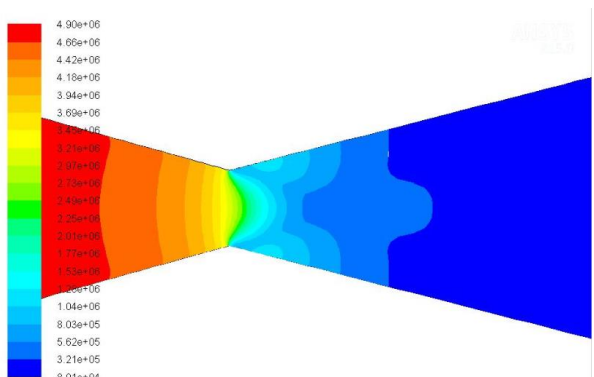 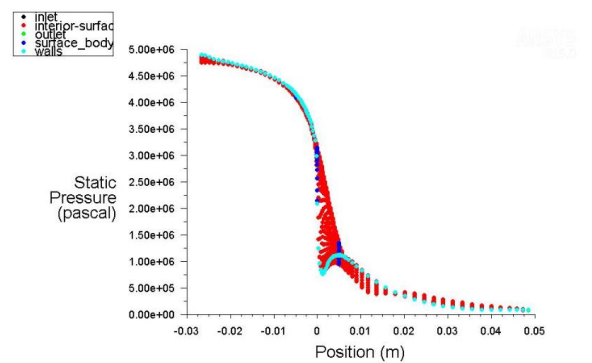

   Fig.8: Static pressure contour .                                  Fig.9 Static pressure variation along nozzle.

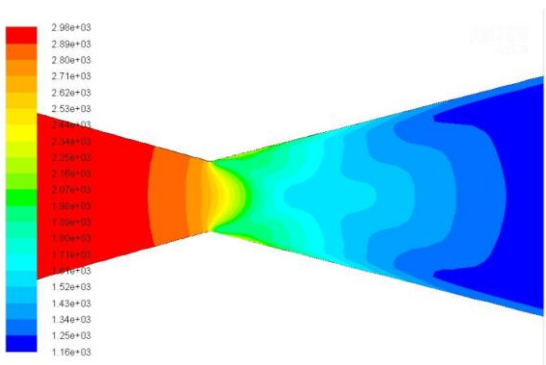 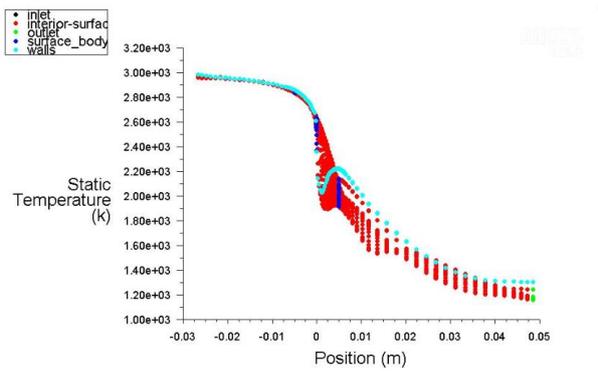

| Fig.10 static temperature contour. | Fig.11 Static temperature variation along nozzle. |

Above graphs shows the flow in the convergent divergent nozzle. There is decrease in the pressure and temperature of gas and increase in the mach number as predicted by threotical calculation with acceptable error.

**5.2 Deorbit simulation:**

The de-orbiting operation the studsat-II is done in the AGI STK tool in the real space time scenario. In the STK, it helps us to model the perturbation effect caused by the non-elliptical, atmospheric drag model and solar radiation pressure, etc. To model these effects, a perturbation model is developed which is 2-body, J2, SGP4, HPOP orbit propagator model. For our scenario, SGP4 model is used for the accurate modeling of the orbit of the satellite by using the 2-line elements which is encoded by the orbital parameter of the satellite body. The HPOP propagator uses Grace gravity Model(GGM03S), SW model, EOP model, SOLRESAP model, and SOLFSMY model to provide the Earth gravity field, Space weather data, Earth orientation parameters, Geomagnetic storm indices, and Solar storm indices space environment in the simulation setup.

The schematic diagram of stk simulation interface is shown in the figure 12 which represents the de-orbit lines of the STUDSAT-II. A Hohmann's orbit transfer is used for the simulation which uses the ΔV1 and ΔV2 burns of the chemical thruster to move it from the orbit of 600 km to the 200 km height. After that, the satellite bodies will itself burn due to the high effect of the air drag at low atmosphere.

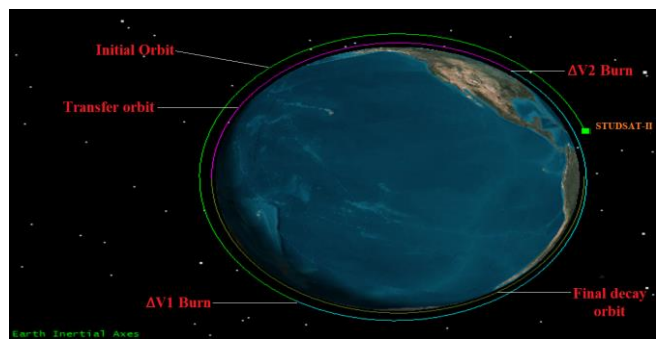

Fig.12: STK simulation interface showing the STUDSAT-II de-orbiting phase.

The simulation results to de-orbiting the STUDSAT-II are shown below. The orbital parameter variations of de-orbiting phase are shown in figure 13-17. Figure 13 represent the semi-major axis variation across the UTC time period. From figure, we can see that, initial semi-major axis is 6978 km, and when ΔV1 is applied through the thruster, there is shift in the height to 6678 km approximate then after sometimes ΔV2 is applied and it moved to 6578 km of the height which is the altitude of the 200 km approximate. The overall process took about 7hours approximate. Similarly, figure 14 represents the variation of inclination across the time which we can see that it continuously changing during de-orbiting phase. The variation taking place is the sinusoidal form due to the 2 burn caused by the thruster. The variation of the true anomaly can be seen in the figure 15. It always varies from 0-360 degree but at the point of the ΔV1 and ΔV2, it causing the spike change in true anomaly variation.

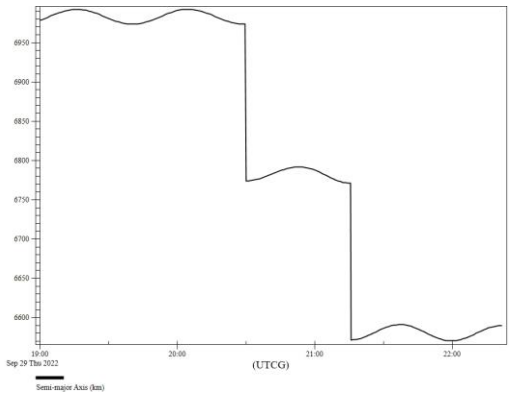

Fig 13: Variation of semi-major axis across time of time of de-orbiting phase.

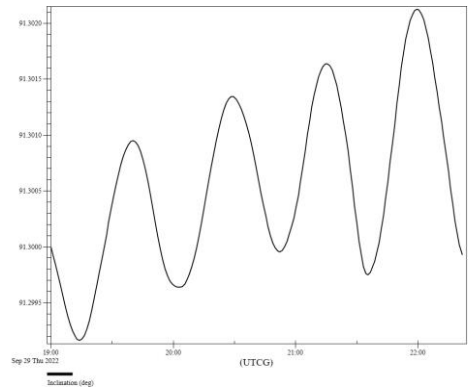

Fig14: Variation of Inclination across de-orbiting phase.

Figure 16 represents the graphical representation of the RAAN across the time period of the de-orbiting, there is slightly change in the RAAN distribution which is about 0.022 degree angle. Similarly, the variation the argument of perigee across the simulation time can be shown in figure 17. Its varies from 0-360 degree across the de-orbiting operation and forming spike change at the point he burns which is mostly dependent to semi-major axis parameter

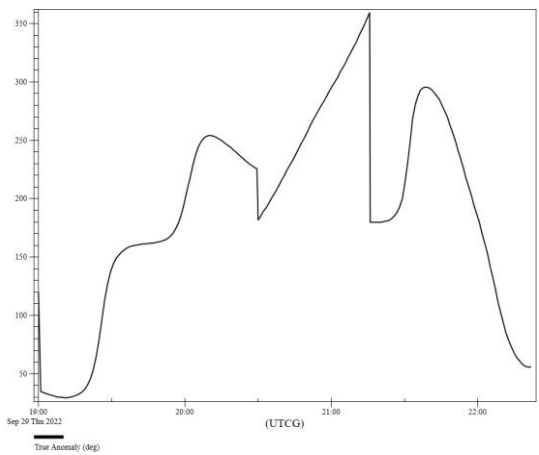

Fig15: Variation of True Anomaly across time of de-orbiting phase.

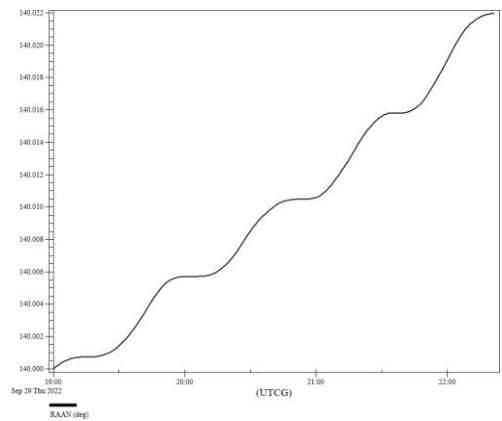

Fig16: Variation of RAAN across time of de-orbiting phase.

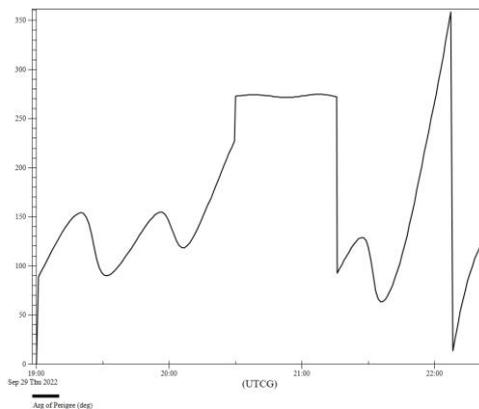

Fig 17: Variation of Argument of Perigee axis across time of during de-orbiting phase.

## 6. CONCLUSIONS

Solid Motor propellant is one of the most favorable solution for the deorbit of satellite due to its simplicity, compactness and stability but there are many challenges. Stability of satellite during deorbit is one of the greatest challenge, as orientation of satellite is plays crucial role in determination of the deorbit paths. Attitude control system should be able to take over small disturbances. High pressure during combustion increases combustion stability but increases system mass. Further research is required to increase the stability of solid propellant. Another challenge is the accuracy of sensor for firing of thruster. Mainly solid motors have been used for the recovery of on-board equipments, attention should be given to on research of deorbiting of satellite at the end-of-life.

**ACKNOWLEDGMENTS**

I would like to express my sincere gratitude to Mr. Sandesh Hegde for supervising in this work. Without his guidance this work would not have been completed. I would like to thank entire family of Centre for Small Satellite Research for the completion of this work.